\PassOptionsToPackage{unicode}{hyperref}
\PassOptionsToPackage{hyphens}{url}
\documentclass[
  11pt,
]{article}
\usepackage{amsmath,amssymb}
\usepackage{fontspec}
\usepackage{iftex}
\iffalse

\else 
  \usepackage{unicode-math} 
  \defaultfontfeatures{Scale=MatchLowercase}
  \defaultfontfeatures[\rmfamily]{Ligatures=TeX,Scale=1}
\fi
\iffalse\else
\fi
\IfFileExists{upquote.sty}{\usepackage{upquote}}{}
\makeatletter
\@ifundefined{KOMAClassName}{
  \IfFileExists{parskip.sty}{%
    \usepackage{parskip}
  }{
    \setlength{\parindent}{0pt}
    \setlength{\parskip}{6pt plus 2pt minus 1pt}}
}{
  \KOMAoptions{parskip=half}}
\makeatother
\usepackage{xcolor}
\usepackage[margin=1in]{geometry}
\usepackage{color}
\usepackage{fancyvrb}

\DefineVerbatimEnvironment{Highlighting}{Verbatim}{commandchars=\\\{\}}
\newenvironment{Shaded}{}{}
\newcommand{\AlertTok}[1]{\textcolor[rgb]{1.00,0.00,0.00}{\textbf{#1}}}

\newcommand{\AttributeTok}[1]{\textcolor[rgb]{0.49,0.56,0.16}{#1}}

\newcommand{\CommentTok}[1]{\textcolor[rgb]{0.38,0.63,0.69}{\textit{#1}}}

\newcommand{\ConstantTok}[1]{\textcolor[rgb]{0.53,0.00,0.00}{#1}}

\newcommand{\DecValTok}[1]{\textcolor[rgb]{0.25,0.63,0.44}{#1}}

\newcommand{\FloatTok}[1]{\textcolor[rgb]{0.25,0.63,0.44}{#1}}
\newcommand{\FunctionTok}[1]{\textcolor[rgb]{0.02,0.16,0.49}{#1}}

\newcommand{\NormalTok}[1]{#1}

\newcommand{\OtherTok}[1]{\textcolor[rgb]{0.00,0.44,0.13}{#1}}

\newcommand{\SpecialCharTok}[1]{\textcolor[rgb]{0.25,0.44,0.63}{#1}}

\newcommand{\StringTok}[1]{\textcolor[rgb]{0.25,0.44,0.63}{#1}}

\usepackage{longtable,booktabs,array}
\usepackage{calc} 
\usepackage{etoolbox}
\makeatletter
\patchcmd\longtable{\par}{\if@noskipsec\mbox{}\fi\par}{}{}
\makeatother
\IfFileExists{footnotehyper.sty}{\usepackage{footnotehyper}}{\usepackage{footnote}}
\makesavenoteenv{longtable}
\providecommand{\tightlist}{%
  \setlength{\itemsep}{0pt}\setlength{\parskip}{0pt}}
\usepackage{amsmath}
\usepackage{amssymb}
\usepackage{booktabs}
\usepackage{longtable}
\usepackage{array}
\usepackage{hyperref}
\hypersetup{colorlinks=true, linkcolor=blue, urlcolor=blue, citecolor=blue}
\usepackage{setspace}
\ifLuaTeX
  \usepackage{selnolig}  
\fi
\IfFileExists{bookmark.sty}{\usepackage{bookmark}}{\usepackage{hyperref}}
\IfFileExists{xurl.sty}{\usepackage{xurl}}{} 
\hypersetup{
  pdftitle={ICCDesign: An R Package for the Design and Analysis of ICC-Based Reliability Studies with Continuous Responses},
  pdfauthor={Yundan Zhang;Ziyu Liu; Ruilin Ma; Chenge Gao},
  pdfkeywords={ICC; intraclass correlation coefficient; reliability
analysis; sample size calculation; R package; Shiny application},
  hidelinks,
  pdfcreator={LaTeX via pandoc}}

\title{ICCDesign: An R Package for the Design and Analysis of ICC-Based
Reliability Studies with Continuous Responses}
\author{Ziyu Liu\thanks{Corresponding author. School of Statistics, East China Normal University, Shanghai, China. Email: 1755454769@qq.com} \and Yundan Zhang \and Ruilin Ma \and Chenge Gao }
\date{School of Statistics, East China Normal University, Shanghai,
China}

\begin{document}
\maketitle
\begin{abstract}
The intraclass correlation coefficient (ICC) is among the most widely
used statistics in reliability research, playing a central role in
medical measurement, psychological assessment, and behavioral science.
However, practical application of ICC faces two major obstacles. First,
ICC can be organized into multiple forms under the McGraw and Wong
(1996) framework---including six widely reported standard forms and four
additional design combinations---and researchers must select the
appropriate form based on their study design, yet existing guidelines
are not always operationalized in software interfaces. Second, available
R tools are highly fragmented: sample size calculation, ICC estimation
with confidence intervals, and reliability evaluation are distributed
across separate packages, compelling researchers to switch between tools
and increasing the risk of analytical errors. This paper introduces the
\textbf{ICCDesign} package, designed specifically to provide an
integrated workflow for ICC-based reliability studies with continuous
responses, assuming one continuous rating per subject--rater cell. The
package integrates four core functionalities: (1) point estimation,
ANOVA-based confidence intervals, and implemented hypothesis tests for
the ten supported ICC design settings following the McGraw and Wong
(1996) framework, with a built-in four-step decision framework guiding
users toward an appropriate ICC form; (2) sample size planning based on
Zou's (2012) closed-form formulas, supporting two planning modes and an
inverse assurance calculation; (3) automated reliability evaluation
based on Koo and Li (2016) criteria, with an uncertainty notification
when the confidence interval spans the 0.75 good-reliability threshold;
and (4) an interactive Shiny web application covering the main analysis
and planning functionalities. ICCDesign is available from GitHub at
https://github.com/KlariZhang/ICCDesign.
\end{abstract}

\hypertarget{introduction}{%
\subsection{1. Introduction}\label{introduction}}

In repeated measurement studies, inter-rater agreement studies, and
measurement instrument validation, researchers require a statistic that
can quantify reliability among repeated measurements on the same object.
The intraclass correlation coefficient (ICC) serves this purpose: built
upon a variance components model, it partitions observed variability
into subject-related and error-related components---with additional
rater-related components in two-way designs. Depending on the selected
form, ICC can quantify either absolute agreement or relative consistency
among raters. ICC has been widely applied to evaluate measurement
agreement in medical imaging, assess inter-rater reliability of clinical
scales, and analyze test-retest reliability of psychological instruments
(Shrout \& Fleiss, 1979; McGraw \& Wong, 1996; Koo \& Li, 2016).

Despite its widespread use, the correct application of ICC presents
substantial challenges in practice. Shrout and Fleiss (1979) first
systematically categorized ICC into three model forms, and McGraw and
Wong (1996) subsequently organized these into ten design combinations,
each corresponding to different research design assumptions. Selecting
the appropriate ICC form requires researchers to address four core
questions: whether the same set of raters evaluates all subjects;
whether rater effects are treated as random or fixed; whether the unit
of analysis is a single rating or the mean of multiple ratings; and
whether the target is absolute agreement or relative consistency.
Different answers correspond to different statistical models and
computation formulas, and incorrect selection may lead to misleading
reliability interpretations.

It should be noted that ICCDesign does not propose a new ICC estimator;
it consolidates established ICC estimation, inference, planning, and
interpretation procedures into a unified interface. Within the R
software ecosystem, existing ICC-related tools are distributed in a
fragmented manner. The \texttt{irr} package (Gamer et al., 2019) and the
\texttt{psych} package (Revelle, 2024) provide point estimates and
confidence intervals for ICC, but cover only a subset of ICC forms and
offer no guidance for form selection. The \texttt{ICC.Sample.Size}
package (Wolak et al., 2012), based on the method of Zou (2012),
provides sample size calculation but supports only a single calculation
mode and contains no ICC estimation functionality. The \texttt{irrNA}
package (Brueckl, 2022) extends \texttt{irr} to handle randomly
incomplete datasets but likewise lacks design-stage functionality. This
fragmentation means that conducting a complete ICC reliability study
requires researchers to coordinate at least two or three separate
packages, while inconsistent interfaces across packages increase the
risk of operational errors.

The \textbf{ICCDesign} package introduced in this paper aims to fill
these gaps by providing an integrated solution for ICC-based reliability
studies with continuous responses. The main contributions of the package
are as follows:

\begin{itemize}
\tightlist
\item
  Implementation of point estimation, ANOVA-based confidence intervals,
  and hypothesis testing for the ten supported ICC design combinations
  under the McGraw and Wong (1996) framework, covering six standard
  forms and four additional combinations;
\item
  A built-in four-step decision framework that guides users from study
  design specifications toward an appropriate ICC form through
  structured parameter settings, with explicit user feedback for special
  cases (automatic mapping and use-with-caution combinations);
\item
  Sample size planning based on Zou's (2012) closed-form formulas,
  together with inverse assurance calculations;
\item
  Automated reliability grading integrated with the Koo and Li (2016)
  criteria, with automatic generation of standardized text-format
  analysis reports;
\item
  An interactive Shiny web application reducing the need to write R code
  once the package is installed.
\end{itemize}

The remainder of this paper is organized as follows. Section 2 reviews
the statistical foundations of ICC and describes the methods implemented
in ICCDesign. Section 3 describes the package structure and core
functionalities in detail. Section 4 presents a systematic comparison of
ICCDesign with existing R packages. Section 5 demonstrates the use of
the package through concrete examples. Section 6 provides discussion and
directions for future development.

\begin{center}\rule{0.5\linewidth}{0.5pt}\end{center}

\hypertarget{methodological-background}{%
\subsection{2. Methodological
Background}\label{methodological-background}}

\hypertarget{anova-framework-and-three-statistical-models}{%
\subsubsection{2.1 ANOVA Framework and Three Statistical
Models}\label{anova-framework-and-three-statistical-models}}

The statistical definition of ICC is built upon the analysis of variance
(ANOVA) framework. Let \(n\) denote the number of subjects and \(k\) the
number of raters, with \(Y_{ij}\) representing the continuous rating
assigned to subject \(i\) by rater \(j\). The current implementation
assumes one rating per subject--rater cell and a balanced data matrix
after preprocessing. ICCDesign implements three statistical models
corresponding to different research design settings.

The following mean squares are used throughout: MSR (subject mean
square), MSC (rater/column mean square), MSE (residual/error mean
square), and MSW (within-group mean square for the one-way model).

\textbf{One-way random effects model.} This model is used when raters
are not treated as a fixed common set across all subjects, corresponding
to ICC(1,x) types. This corresponds to designs in which rater effects
cannot be separated as a crossed factor:

\[Y_{ij} = \mu + \alpha_i + \varepsilon_{ij}\]

where \(\mu\) is the overall mean,
\(\alpha_i \sim \mathcal{N}(0, \sigma_\alpha^2)\) is the subject random
effect, and
\(\varepsilon_{ij} \sim \mathcal{N}(0, \sigma_\varepsilon^2)\) is the
residual error. The expected mean squares are:

\[E(\text{MSR}) = \sigma_\varepsilon^2 + k\sigma_\alpha^2, \quad E(\text{MSW}) = \sigma_\varepsilon^2\]

\textbf{Two-way random effects model.} This model is used when all
subjects are rated by the same set of raters who are treated as randomly
sampled from a broader rater population, corresponding to ICC(2,x)
types. Since each subject--rater cell contains only one observation, the
subject-by-rater interaction and pure measurement error cannot be
separated. The combined residual term \(e_{ij}\) absorbs both sources:

\[Y_{ij} = \mu + \alpha_i + \beta_j + e_{ij}\]

where \(\beta_j \sim \mathcal{N}(0, \sigma_\beta^2)\) is the rater
random effect and \(e_{ij}\) represents the combined interaction and
measurement error. The expected mean squares are:

\[E(\text{MSR}) = \sigma_e^2 + k\sigma_\alpha^2, \quad E(\text{MSC}) = \sigma_e^2 + n\sigma_\beta^2, \quad E(\text{MSE}) = \sigma_e^2\]

where \(\sigma_e^2\) denotes the combined residual variance. Note that
\(E(\text{MSC})\) applies to the two-way random effects model only; in
the two-way mixed effects model, rater effects are fixed and
\(\sigma_\beta^2\) is not defined.

\textbf{Two-way mixed effects model.} This model is used when the raters
constitute a specific fixed group, corresponding to ICC(3,x) types:

\[Y_{ij} = \mu + \alpha_i + \beta_j + e_{ij}\]

where \(\beta_j\) are fixed effects satisfying \(\sum_j \beta_j = 0\)
and \(e_{ij}\) is the combined residual. MSC is not used for standard
ICC(3,x) consistency calculations but enters the supplementary
absolute-agreement calculations (see Section 2.2). The expected mean
squares for subject and residual terms are:

\[E(\text{MSR}) = \sigma_e^2 + k\sigma_\alpha^2, \quad E(\text{MSE}) = \sigma_e^2\]

\textbf{Statistical meaning of the \texttt{interaction} parameter.} The
\texttt{interaction} parameter controls how the two-way ANOVA residual
is computed. When \texttt{interaction\ =\ TRUE} (default), the residual
sum of squares is computed as
\(\text{SS}_e = \sum_{ij}(Y_{ij} - \bar{Y}_{i\cdot} - \bar{Y}_{\cdot j} + \bar{Y})^2\)
with degrees of freedom \((n-1)(k-1)\). When
\texttt{interaction\ =\ FALSE}, ICCDesign uses a pooled residual
calculation by merging the rater sum of squares into the residual:
\(\text{SS}_e \leftarrow \text{SS}_e + \text{SS}_c\), expanding the
residual degrees of freedom to \((n-1)(k-1) + (k-1)\). This is an
optional computational setting and should not be interpreted as a fully
identified no-interaction model with replicated cells. The default
\texttt{interaction\ =\ TRUE} is recommended for standard ICC analyses.

\hypertarget{classification-of-icc-forms}{%
\subsubsection{2.2 Classification of ICC
Forms}\label{classification-of-icc-forms}}

Shrout and Fleiss (1979) first systematically categorized ICC into three
model forms. McGraw and Wong (1996) organized this into a set of ten
design combinations along two additional dimensions. ICCDesign exposes
these ten combinations, classifying them as six standard types (the most
commonly reported in the literature) and four additional combinations;
this classification reflects the package interface rather than a
separate literature-based taxonomy.

\textbf{Dimension 1: Unit of Measurement.} When the reliability of a
single rating is of interest, the single-measure ICC is used; when the
reliability of the mean of \(k\) ratings is of interest, the
average-measure ICC is used. The average-measure ICC refers specifically
to the reliability of the mean of the \(k\) raters included in the study
design. The latter generally yields higher values due to error reduction
through averaging.

\textbf{Dimension 2: Type of Consistency.} Absolute agreement treats
systematic rater differences as a source of error, requiring ratings to
agree numerically across raters; consistency allows for systematic rater
mean differences and requires only that raters maintain consistent
relative patterns across subjects.

\textbf{Use-with-caution combination 1 (automatic mapping).} The
consistency ICC under the two-way random effects model (ICC(2,x)\_C) and
the consistency ICC under the two-way mixed effects model (ICC(3,x))
share identical ANOVA-based point estimators and inference formulas,
although their design interpretations differ. ICCDesign uses the same
computational routine while retaining the user-specified ICC(2,x)\_C
label in the output (e.g., \texttt{icc\_code\ =\ "2,1,consistency"}),
and issues a NOTE-level notification via \texttt{message()} to inform
users of this mapping behavior.

\textbf{Use-with-caution combination 2 (fixed-rater absolute
agreement).} The absolute agreement ICC under the two-way mixed effects
model (ICC(3,x)\_A) should be used with caution; inference is restricted
to the specific fixed raters in the study and does not generalize to a
broader rater population. ICCDesign provides computational support for
this combination but issues a WARNING-level alert via
\texttt{warning()}, recommending that users use ICC(2,x) instead.

The ten design combinations are summarized in Table 1.

\textbf{Table 1: Ten ICC Design Combinations}

\begin{longtable}[]{@{}
  >{\raggedright\arraybackslash}p{(\columnwidth - 10\tabcolsep) * \real{0.1163}}
  >{\raggedright\arraybackslash}p{(\columnwidth - 10\tabcolsep) * \real{0.0814}}
  >{\raggedright\arraybackslash}p{(\columnwidth - 10\tabcolsep) * \real{0.1512}}
  >{\raggedright\arraybackslash}p{(\columnwidth - 10\tabcolsep) * \real{0.2093}}
  >{\raggedright\arraybackslash}p{(\columnwidth - 10\tabcolsep) * \real{0.1977}}
  >{\raggedright\arraybackslash}p{(\columnwidth - 10\tabcolsep) * \real{0.2442}}@{}}
\toprule\noalign{}
\begin{minipage}[b]{\linewidth}\raggedright
ICC Form
\end{minipage} & \begin{minipage}[b]{\linewidth}\raggedright
Model
\end{minipage} & \begin{minipage}[b]{\linewidth}\raggedright
Rating Unit
\end{minipage} & \begin{minipage}[b]{\linewidth}\raggedright
Consistency Type
\end{minipage} & \begin{minipage}[b]{\linewidth}\raggedright
Interface Status
\end{minipage} & \begin{minipage}[b]{\linewidth}\raggedright
Typical Application
\end{minipage} \\
\midrule\noalign{}
\endhead
\bottomrule\noalign{}
\endlastfoot
ICC(1,1) & One-way random & Single & Absolute & Standard & Raters not
common across all subjects; single-rating reliability \\
ICC(1,k) & One-way random & Average & Absolute & Standard & Raters not
common across all subjects; mean-rating reliability \\
ICC(2,1) & Two-way random & Single & Absolute & Standard & Common raters
sampled from a broader population; systematic differences are of
concern \\
ICC(2,k) & Two-way random & Average & Absolute & Standard & Common
raters sampled from a broader population; mean-rating reliability \\
ICC(3,1) & Two-way mixed & Single & Consistency & Standard & Fixed
raters; systematic rater mean differences are not counted as
disagreement \\
ICC(3,k) & Two-way mixed & Average & Consistency & Standard & Fixed
raters; systematic rater mean differences are not counted as
disagreement; mean-rating reliability \\
ICC(2,1)\_C & Two-way random & Single & Consistency & Additional / use
with caution & Computational routine shared with ICC(3,1); output label
retained; NOTE issued \\
ICC(2,k)\_C & Two-way random & Average & Consistency & Additional / use
with caution & Computational routine shared with ICC(3,k); output label
retained; NOTE issued \\
ICC(3,1)\_A & Two-way mixed & Single & Absolute & Additional / use with
caution & Inference restricted to fixed raters; WARNING issued \\
ICC(3,k)\_A & Two-way mixed & Average & Absolute & Additional / use with
caution & Inference restricted to fixed raters; WARNING issued \\
\end{longtable}

\emph{Note: ``Additional / use with caution'' means the combination is
exposed by ICCDesign beyond the six commonly reported standard forms.
Users should be aware of the design interpretation limitations described
in Section 2.2.}

\hypertarget{point-estimation-formulas}{%
\subsubsection{2.3 Point Estimation
Formulas}\label{point-estimation-formulas}}

The following formulas correspond to those currently implemented in
\texttt{icc\_tool\_point()} for the supported forms, adapted from the
McGraw and Wong (1996) framework. ANOVA-based point estimates may
occasionally be negative when between-group variability is low;
ICCDesign reports the computed value without truncation at the point
estimation stage. The mean squares MSR, MSC, MSE, and MSW are returned
in \texttt{\$anova\_summary} for inspection.

\textbf{One-way random effects model:}

\[\widehat{\text{ICC}}(1,1) = \frac{\text{MSR} - \text{MSW}}{\text{MSR} + (k-1)\text{MSW}}, \quad \widehat{\text{ICC}}(1,k) = \frac{\text{MSR} - \text{MSW}}{\text{MSR}}\]

\textbf{Two-way random effects model --- absolute agreement:}

\[\widehat{\text{ICC}}(2,1) = \frac{\text{MSR} - \text{MSE}}{\text{MSR} + (k-1)\text{MSE} + \frac{k}{n}(\text{MSC} - \text{MSE})}, \quad \widehat{\text{ICC}}(2,k) = \frac{\text{MSR} - \text{MSE}}{\text{MSR} + \frac{1}{n}(\text{MSC} - \text{MSE})}\]

\textbf{Two-way mixed effects model --- consistency} (computational
routine shared with ICC(2,x)\_C):

\[\widehat{\text{ICC}}(3,1) = \frac{\text{MSR} - \text{MSE}}{\text{MSR} + (k-1)\text{MSE}}, \quad \widehat{\text{ICC}}(3,k) = \frac{\text{MSR} - \text{MSE}}{\text{MSR}}\]

The supplementary types ICC(3,x)\_A use the same implemented formulas as
ICC(2,x); although the point-estimate formula is shared, the inferential
target differs.

\hypertarget{anova-based-f-confidence-intervals-and-satterthwaite-type-approximation}{%
\subsubsection{2.4 ANOVA-Based F Confidence Intervals and
Satterthwaite-Type
Approximation}\label{anova-based-f-confidence-intervals-and-satterthwaite-type-approximation}}

ICCDesign computes F-distribution-based confidence intervals via
\texttt{icc\_tool\_ci()}, following McGraw and Wong (1996) Table 7 and
assuming balanced data. All reported user-facing bounds are truncated to
the interpretable reliability range \([0, 1]\); the analytical bounds
before truncation are not separately returned. The formulas below use
\(F_p(df_1, df_2)\) to denote the \(p\)-th quantile of the F
distribution with degrees of freedom \(df_1\) and \(df_2\). The
confidence intervals are sensitive to the normality and balanced-design
assumptions underlying the ANOVA decomposition.

\textbf{Type 1 intervals} (applicable to ICC(1,x), ICC(3,x),
ICC(2,x)\_C): based directly on the standard F distribution without
correction.

For ICC(1,1), the \((1-\alpha)\) interval is:

\[\left[\frac{F_0/F_{1-\alpha/2} - 1}{F_0/F_{1-\alpha/2} + (k-1)},\quad \frac{F_0/F_{\alpha/2} - 1}{F_0/F_{\alpha/2} + (k-1)}\right]\]

where \(F_0 = \text{MSR}/\text{MSW}\),
\(F_{1-\alpha/2} = F_{1-\alpha/2}(n-1,\ n(k-1))\), and
\(F_{\alpha/2} = F_{\alpha/2}(n-1,\ n(k-1))\). Dividing by the larger
quantile \(F_{1-\alpha/2}\) yields the lower bound; dividing by the
smaller quantile \(F_{\alpha/2}\) yields the upper bound.

For ICC(1,k), the interval takes a different form. Since
\(\widehat{\text{ICC}}(1,k) = 1 - \text{MSW}/\text{MSR}\), the
confidence interval is derived directly from the F quantiles without
dividing by an observed F ratio:

\[\left[1 - \frac{1}{F_{1-\alpha/2}(n-1,\ n(k-1))},\quad 1 - \frac{1}{F_{\alpha/2}(n-1,\ n(k-1))}\right]\]

For ICC(3,1) and ICC(2,1)\_C, the same form as ICC(1,1) applies with
\(F_0 = \text{MSR}/\text{MSE}\) and degrees of freedom
\((n-1,\ (n-1)(k-1))\). For ICC(3,k) and ICC(2,k)\_C, the same structure
as ICC(1,k) applies with degrees of freedom \((n-1,\ (n-1)(k-1))\):

\[\left[1 - \frac{1}{F_{1-\alpha/2}(n-1,\ (n-1)(k-1))},\quad 1 - \frac{1}{F_{\alpha/2}(n-1,\ (n-1)(k-1))}\right]\]

\textbf{Type 2 intervals} (applicable to ICC(2,x)\_A and ICC(3,x)\_A):
since the absolute agreement ICC under two-way models is not a monotone
function of a simple F ratio, a Satterthwaite-type approximation is
required to obtain corrected degrees of freedom. Define:

\[a = \frac{k \cdot \text{MSC}}{n} + (n-1)\cdot\text{MSE}, \quad b = \frac{(n-1) \cdot k \cdot \text{MSE}}{n}\]

\[df_{\text{corr}} = \frac{(a + b)^2}{a^2/(n-1) + b^2/((n-1)(k-1))}\]

For ICC(2,1) and ICC(3,1)\_A, let
\(F_{\text{val}} = \text{MSR}/\text{MSE}\) and define the correction
factor \(\Delta = (k-1) + k(\text{MSC}-\text{MSE})/(n\cdot\text{MSE})\).
The confidence interval is:

\[\left[\frac{F_{\text{val}}/F_{1-\alpha/2}^* - 1}{F_{\text{val}}/F_{1-\alpha/2}^* + \Delta},\quad \frac{F_{\text{val}}/F_{\alpha/2}^* - 1}{F_{\text{val}}/F_{\alpha/2}^* + \Delta}\right]\]

where \(F_{1-\alpha/2}^* = F_{1-\alpha/2}(n-1, df_{\text{corr}})\) and
\(F_{\alpha/2}^* = F_{\alpha/2}(n-1, df_{\text{corr}})\).

For average-measure two-way absolute-agreement ICCs (ICC(2,k) and
ICC(3,k)\_A), ICCDesign applies the corresponding Satterthwaite-type
interval following the McGraw and Wong (1996) framework; the
algebraically lengthy expression is omitted here and implemented in
\texttt{icc\_tool\_ci()}.

For ICC(3,x)\_A, the same computational formula is provided as a
use-with-caution option; its interpretation is restricted to the fixed
raters included in the study.

\hypertarget{hypothesis-testing}{%
\subsubsection{2.5 Hypothesis Testing}\label{hypothesis-testing}}

ICCDesign implements one-tailed F-tests (\(H_1: \text{ICC} > \rho_0\))
via \texttt{icc\_calc\_f\_test()} following McGraw and Wong (1996) Table
8, supporting only the upper-tail direction. Two categories of tests are
supported:

\textbf{Default zero-ICC test} (always executed):
\(H_0: \text{ICC} = 0\). Results are stored in
\texttt{\$icc\_result\$F\_test\_null}. The F statistics and degrees of
freedom are:

\begin{itemize}
\tightlist
\item
  ICC(1,x): \(F = \text{MSR}/\text{MSW}\), df \((n-1,\ n(k-1))\)
\item
  ICC(3,x), ICC(2,x)\_C: \(F = \text{MSR}/\text{MSE}\), df
  \((n-1,\ (n-1)(k-1))\)
\item
  ICC(2,x)\_A: \(F = \text{MSR}/\text{MSE}\), df \((n-1,\ (n-1)(k-1))\)
\end{itemize}

\textbf{Optional threshold test} (conditionally executed):
\(H_0: \text{ICC} = \rho_0\), where \(\rho_0 \in [0, 1)\) is a
user-specified minimum acceptable reliability threshold. This test is
executed only when the user provides the \texttt{rho0} argument; results
are stored in \texttt{\$icc\_result\$F\_test\_rho0} (which is
\texttt{NULL} when \texttt{rho0} is not provided).

ICCDesign implements the following threshold-test F statistics for the
supported forms in the current software implementation; for both single-
and average-measure variants within each model type, the same
F-statistic formula is used.

For ICC(1,x):

\[F = \frac{(\text{MSR}/\text{MSW}) \cdot (1 - \rho_0)}{1 + (k-1)\rho_0}\]

For ICC(3,x) and ICC(2,x)\_C:

\[F = \frac{(\text{MSR}/\text{MSE}) \cdot (1 - \rho_0)}{1 + (k-1)\rho_0}\]

For ICC(2,x)\_A, the non-zero F statistic additionally incorporates the
rater variance component:

\[F = \frac{\text{MSR}(1 - \rho_0)}{\text{MSE}(1 + (k-1)\rho_0) + \frac{k(\text{MSC} - \text{MSE})\rho_0}{n}}\]

All p-values are one-tailed (upper tail). The current implementation
does not support two-sided or lower-tail tests.

\hypertarget{sample-size-calculation-zou-2012-closed-form-formulas}{%
\subsubsection{2.6 Sample Size Calculation: Zou (2012) Closed-form
Formulas}\label{sample-size-calculation-zou-2012-closed-form-formulas}}

ICCDesign implements two planning formulas and an inverse assurance
calculation based on Zou (2012), controlled by \texttt{assurance} =
\(\gamma\) (assurance probability, default 0.8, satisfying
\(0 < \gamma < 1\)) and \texttt{alpha} = \(\alpha\) (significance level,
default 0.05). Here \(n\) denotes the number of subjects; the total
number of ratings in a balanced design is \(n \times k\), where \(k\) is
the number of raters and is treated as fixed. Sample size results from
Modes 1 and 2 are returned as integers via \texttt{ceiling()}. Note that
the sample size formulas depend on \(\rho\), \(k\), and the planning
criterion, but not on the specific ICC type; the four-step design
parameters are accepted for interface consistency and input validation
only.

\textbf{Mode 1: Lower bound assurance} (\texttt{method\ =\ "lower"}).
Finds the minimum number of subjects \(n\) satisfying
\(P(\hat{L} \geq \rho_0) \geq \gamma\), where \(\hat{L}\) is the lower
bound of the ICC confidence interval, \(\rho\) is the anticipated true
ICC, and \(\rho_0\) is the acceptable minimum. This mode requires
\(\rho > \rho_0\); the function raises an error if this condition is not
met. Parameters \(0 < \alpha < 1\), \(0 < \gamma < 1\), and \(k \geq 2\)
are also required.

\[N = 1 + \frac{2(z_{1-\alpha} + z_{\gamma})^2 k}{(k-1)\left[\ln\left(\dfrac{F_\rho}{F_{\rho_0}}\right)\right]^2}\]

where \(F_\rho = \dfrac{1 + (k-1)\rho}{1 - \rho}\),
\(F_{\rho_0} = \dfrac{1 + (k-1)\rho_0}{1 - \rho_0}\),
\(z_{1-\alpha} = \Phi^{-1}(1-\alpha)\) (one-tailed quantile),
\(z_{\gamma} = \Phi^{-1}(\gamma)\).

\textbf{Mode 2: Confidence interval half-width}
(\texttt{method\ =\ "width"}). The parameter \texttt{omega} =
\(\omega > 0\) specifies the desired half-width of the confidence
interval. Finds the minimum \(n\) satisfying
\(P(\hat{\omega} \leq \omega) \geq \gamma\), where \(\hat{\omega}\) is
the realized CI half-width:

\[N = 1 + \left[\frac{Az_{1-\alpha/2} + \sqrt{A^2 z_{1-\alpha/2}^2 + 4\omega z_{1-\alpha/2} z_{\gamma} A|B|}}{\omega\sqrt{2k(k-1)}}\right]^2\]

where \(A = (1-\rho)(1+(k-1)\rho)\), \(B = k - 2 + 2\rho - 2k\rho\), and
\(z_{1-\alpha/2} = \Phi^{-1}(1-\alpha/2)\) (two-tailed quantile,
distinct from the one-tailed \(z_{1-\alpha}\) in Mode 1). The absolute
value \(|B|\) arises from the Zou (2012) approximation derivation.

\textbf{Mode 3: Inverse assurance calculation}
(\texttt{method\ =\ "power"}). Given subject count \(n\), computes the
assurance probability. The underlying \texttt{icc\_power()} function
supports \texttt{method\ =\ "lower"} (default) and
\texttt{method\ =\ "width"} criteria. For \texttt{method\ =\ "width"},
the assurance probability is computed as an approximation and should be
treated as a planning aid.

The \texttt{rating\_target} parameter provides semantic specification of
\(\rho_0\): \texttt{"poor"} and \texttt{"moderate"} both map to 0.50
(the Poor--Moderate boundary), \texttt{"good"} maps to 0.75, and
\texttt{"excellent"} maps to 0.90.

\hypertarget{reliability-grading-criteria-koo-li-2016}{%
\subsubsection{2.7 Reliability Grading Criteria (Koo \& Li,
2016)}\label{reliability-grading-criteria-koo-li-2016}}

ICCDesign performs automated reliability grading using the Koo and Li
(2016) thresholds, applied to the \((1-\alpha)\) CI lower bound (rather
than the point estimate). By default (\texttt{alpha\ =\ 0.05}), the 95\%
CI lower bound is used; if the user modifies \texttt{alpha}, ICCDesign
applies the same threshold rule to the resulting CI level. This
extension to other confidence levels is a software convention rather
than a separate recommendation from Koo and Li (2016). The grading is
interpretive and should not replace study-specific decision thresholds.

\textbf{Table 2: Reliability Grading Criteria}

\begin{longtable}[]{@{}ll@{}}
\toprule\noalign{}
\((1-\alpha)\) CI Lower Bound (default: 95\%) & Reliability Grade \\
\midrule\noalign{}
\endhead
\bottomrule\noalign{}
\endlastfoot
\textless{} 0.50 & Poor \\
{[}0.50, 0.75) & Moderate \\
{[}0.75, 0.90) & Good \\
$\geq$ 0.90 & Excellent \\
\end{longtable}

When the CI lower bound falls below 0.75 while the upper bound reaches
or exceeds 0.75, the \texttt{explanation} field automatically appends an
uncertainty notification, recommending that researchers increase the
sample size for a more precise estimate. No analogous notification is
triggered at the 0.50 or 0.90 thresholds.

\begin{center}\rule{0.5\linewidth}{0.5pt}\end{center}

\hypertarget{the-iccdesign-package}{%
\subsection{3. The ICCDesign Package}\label{the-iccdesign-package}}

\hypertarget{package-architecture-and-installation}{%
\subsubsection{3.1 Package Architecture and
Installation}\label{package-architecture-and-installation}}

ICCDesign is developed for R (version $\geq$ 4.1.0), with core dependencies
limited to \texttt{stats} (base R) and \texttt{shiny}, under the
\texttt{GPL\ (\textgreater{}=\ 3)} license, current version 0.1.0. The
package can be installed from GitHub using \texttt{devtools} (Wickham et
al., 2022) or similar tools:

\begin{Shaded}
\begin{Highlighting}[]
\NormalTok{devtools}\SpecialCharTok{::}\FunctionTok{install\_github}\NormalTok{(}\StringTok{"KlariZhang/ICCDesign"}\NormalTok{)}
\FunctionTok{library}\NormalTok{(ICCDesign)}
\end{Highlighting}
\end{Shaded}

The package provides four user-level main functions:

\begin{itemize}
\tightlist
\item
  \texttt{icc\_calc()}: top-level ICC analysis function
\item
  \texttt{icc\_sample\_size()}: unified sample size and power
  calculation interface
\item
  \texttt{run\_icc\_app()}: launch the interactive Shiny application
\item
  \texttt{icc\_preprocess\_data()}: data preprocessing and validation
  utility
\end{itemize}

Internal modules are organized in four layers: a data preprocessing
layer, a design validation and mapping layer, a computation layer (ANOVA
decomposition, point estimation, confidence intervals, F-tests), and an
evaluation and reporting layer.

\hypertarget{data-preprocessing-icc_preprocess_data}{%
\subsubsection{\texorpdfstring{3.2 Data Preprocessing:
\texttt{icc\_preprocess\_data()}}{3.2 Data Preprocessing: icc\_preprocess\_data()}}\label{data-preprocessing-icc_preprocess_data}}

\texttt{icc\_preprocess\_data()} is the dedicated data preprocessing and
validation function among the user-level main functions. It can be
called independently or is automatically invoked internally by
\texttt{icc\_calc()}. The function accepts wide-format data only: rows
represent subjects and columns represent raters; ID columns are not
supported and will cause an error.

Function signature:

\begin{Shaded}
\begin{Highlighting}[]
\FunctionTok{icc\_preprocess\_data}\NormalTok{(data, }\AttributeTok{na.rm =} \ConstantTok{TRUE}\NormalTok{)}
\end{Highlighting}
\end{Shaded}

The function performs three validation and processing steps: (1) format
validation, confirming that the input is a numeric matrix or data frame
(non-numeric values raise an error); (2) dimension validation, requiring
at least 2 rows (subjects) by 2 columns (raters); (3) missing value
handling---\texttt{na.rm\ =\ TRUE} (default) removes rows containing
missing values, which reduces \(n\) and should be noted when
interpreting results; \texttt{na.rm\ =\ FALSE} is primarily diagnostic
and retains missing values with a warning, though complete data are
required for ANOVA-based computation.

The function returns a named list: \texttt{data\_matrix} (standardized
numeric matrix), \texttt{n} (number of valid subjects after row
deletion), \texttt{k} (number of raters), \texttt{warning\_msg} (missing
value warning, \texttt{NULL} if none), and \texttt{error\_msg}
(\texttt{NULL} if data are valid).

\hypertarget{the-four-step-decision-framework-and-icc_calc}{%
\subsubsection{\texorpdfstring{3.3 The Four-Step Decision Framework and
\texttt{icc\_calc()}}{3.3 The Four-Step Decision Framework and icc\_calc()}}\label{the-four-step-decision-framework-and-icc_calc}}

\texttt{icc\_calc()} is the core analytical function of the package,
with the complete function signature:

\begin{Shaded}
\begin{Highlighting}[]
\FunctionTok{icc\_calc}\NormalTok{(data, same\_raters, }\AttributeTok{rater\_effect =} \ConstantTok{NULL}\NormalTok{,}
\NormalTok{         rating\_type, }\AttributeTok{agreement\_type =} \ConstantTok{NULL}\NormalTok{,}
         \AttributeTok{alpha =} \FloatTok{0.05}\NormalTok{, }\AttributeTok{rho0 =} \ConstantTok{NULL}\NormalTok{, }\AttributeTok{interaction =} \ConstantTok{TRUE}\NormalTok{,}
         \AttributeTok{na.rm =} \ConstantTok{TRUE}\NormalTok{, }\AttributeTok{verbose =} \ConstantTok{TRUE}\NormalTok{)}
\end{Highlighting}
\end{Shaded}

The \textbf{four-step decision parameters} translate the ten ICC design
combinations into four structured parameter choices:

\begin{enumerate}
\def\labelenumi{\arabic{enumi}.}
\tightlist
\item
  \textbf{\texttt{same\_raters}} (logical): whether all subjects are
  rated by the same set of raters. \texttt{FALSE} → one-way random
  effects model (note: \texttt{rater\_effect} and
  \texttt{agreement\_type} must not be specified when
  \texttt{same\_raters\ =\ FALSE}; the one-way model is treated as
  absolute agreement in this interface); \texttt{TRUE} → proceed to next
  step.
\item
  \textbf{\texttt{rater\_effect}} (\texttt{"random"} or
  \texttt{"fixed"}): type of rater effect. \texttt{"random"} → two-way
  random effects model; \texttt{"fixed"} → two-way mixed effects model.
\item
  \textbf{\texttt{rating\_type}} (\texttt{"single"} or
  \texttt{"average"}): unit of analysis.
\item
  \textbf{\texttt{agreement\_type}} (\texttt{"absolute"} or
  \texttt{"consistency"}): target consistency type.
\end{enumerate}

The \texttt{rho0} parameter, when provided, must satisfy
\(\rho_0 \in [0, 1)\). The \texttt{interaction} parameter is an advanced
option affecting the ANOVA decomposition (see Section 2.1); the default
\texttt{interaction\ =\ TRUE} is recommended.

The function executes six internal steps: data preprocessing, parameter
validation, ICC mapping, ANOVA decomposition and ICC computation,
reliability evaluation, and text-format report generation. When
\texttt{verbose\ =\ TRUE}, NOTE-level notifications for automatic
mappings are issued via \texttt{message()} and WARNING-level alerts for
use-with-caution combinations via \texttt{warning()}.

The complete output structure (S3 class \texttt{"icc\_result"}) is as
follows:

\begin{verbatim}
$data_summary              # Data preprocessing summary (n, k, missing value info)
$icc_result
  $icc_type                # Full ICC name
  $icc_code                # ICC code (e.g., "2,1")
  $point_est               # Point estimate (untruncated)
  $ci_level                # Confidence level (= 1 - alpha)
  $ci_lower                # CI lower bound (truncated to [0,1])
  $ci_upper                # CI upper bound (truncated to [0,1])
  $F_test_null             # Default zero-ICC test (always present)
    $H0                    # Null hypothesis statement ("ICC = 0")
    $F_stat                # F statistic
    $df1, $df2             # Degrees of freedom
    $p_value               # One-tailed p-value
  $F_test_rho0             # Threshold test (present only when rho0 is non-NULL)
  $anova_summary           # ANOVA decomposition
    $MSR, $MSC, $MSE, $MSW # Mean squares
    $df1 (subject df), $df2 (residual df), $df3 (rater df)
    $n, $k                 # Number of subjects and raters
$evaluation
  $icc_code, $point_est, $ci_lower
  $rating_en               # Reliability grade
  $explanation             # Interpretation text
$report                    # Standardized text-format analysis report
$warning_msg               # WARNING-level message (NULL or character)
$tip_msg                   # NOTE-level message (NULL or character)
\end{verbatim}

Additional report formats may be accessed via advanced options by
calling the internal function \texttt{icc\_generate\_report()} with
\texttt{format\ =\ "markdown"} or \texttt{format\ =\ "html"}.

\hypertarget{sample-size-planning-icc_sample_size}{%
\subsubsection{\texorpdfstring{3.4 Sample Size Planning:
\texttt{icc\_sample\_size()}}{3.4 Sample Size Planning: icc\_sample\_size()}}\label{sample-size-planning-icc_sample_size}}

\texttt{icc\_sample\_size()} is the unified interface, dispatching to
underlying functions via the \texttt{method} parameter:

\begin{Shaded}
\begin{Highlighting}[]
\FunctionTok{icc\_sample\_size}\NormalTok{(}\AttributeTok{method =} \FunctionTok{c}\NormalTok{(}\StringTok{"lower"}\NormalTok{, }\StringTok{"width"}\NormalTok{, }\StringTok{"power"}\NormalTok{), ..., }\AttributeTok{verbose =} \ConstantTok{TRUE}\NormalTok{)}
\end{Highlighting}
\end{Shaded}

\begin{itemize}
\tightlist
\item
  \texttt{method\ =\ "lower"} → \texttt{icc\_sample\_size\_lower()}:
  lower bound assurance mode
\item
  \texttt{method\ =\ "width"} → \texttt{icc\_sample\_size\_width()}: CI
  half-width mode, parameter \texttt{omega} = \(\omega\)
  (\textbf{half-width})
\item
  \texttt{method\ =\ "power"} → \texttt{icc\_power()}: inverse assurance
  calculation
\end{itemize}

All three modes accept the four-step design parameters for interface
consistency and input validation; the sample size and assurance formulas
themselves are not ICC-type-specific. The three underlying functions
\texttt{icc\_sample\_size\_lower()},
\texttt{icc\_sample\_size\_width()}, and \texttt{icc\_power()} are also
independently exported and can be called directly, though the unified
interface is recommended.

The \texttt{rating\_target} parameter provides semantic \(\rho_0\)
specification: \texttt{"poor"} and \texttt{"moderate"} = 0.50 (the
Poor--Moderate boundary), \texttt{"good"} = 0.75, \texttt{"excellent"} =
0.90.

\hypertarget{interactive-shiny-application-run_icc_app}{%
\subsubsection{\texorpdfstring{3.5 Interactive Shiny Application:
\texttt{run\_icc\_app()}}{3.5 Interactive Shiny Application: run\_icc\_app()}}\label{interactive-shiny-application-run_icc_app}}

\begin{Shaded}
\begin{Highlighting}[]
\FunctionTok{run\_icc\_app}\NormalTok{(}\AttributeTok{host =} \StringTok{"127.0.0.1"}\NormalTok{, }\AttributeTok{port =} \ConstantTok{NULL}\NormalTok{,}
            \AttributeTok{launch.browser =} \FunctionTok{interactive}\NormalTok{(), ...)}
\end{Highlighting}
\end{Shaded}

\texttt{port\ =\ NULL} causes Shiny to automatically select an available
port. The application comprises three functional pages:

\textbf{ICC analysis page}: supports three data input methods (built-in
example dataset, CSV/TSV/TXT file upload, or direct paste). The control
panel exposes the four decision parameters, significance level
\texttt{alpha}, missing value handling option \texttt{na.rm}, and the
advanced \texttt{interaction} parameter. Results include the ICC point
estimate, confidence interval, reliability grade, a reliability interval
visualization displaying the point estimate and CI against the grade
bands, F-test results, and ANOVA decomposition table. The analysis
report can be downloaded as a \texttt{.txt} file.

\textbf{Sample size page}: supports switching among the three
calculation modes and displays a power curve (assurance probability as a
function of the number of subjects). Results can be downloaded as a
\texttt{.csv} file.

\textbf{Data and guide page}: provides data preview and diagnostics, a
tabular ICC selection guide, and the Koo and Li (2016) reliability
grading reference table.

\begin{center}\rule{0.5\linewidth}{0.5pt}\end{center}

\hypertarget{comparison-with-existing-r-packages}{%
\subsection{4. Comparison with Existing R
Packages}\label{comparison-with-existing-r-packages}}

Table 3 presents a systematic comparison of ICCDesign with major
ICC-related R packages. ``Partial'' indicates that the package supports
some but not all aspects of the feature as defined in this paper;
footnotes provide details.

\textbf{Table 3: Comparison of ICCDesign with Existing R Packages}

\begin{longtable}[]{@{}
  >{\raggedright\arraybackslash}p{(\columnwidth - 10\tabcolsep) * \real{0.1607}}
  >{\centering\arraybackslash}p{(\columnwidth - 10\tabcolsep) * \real{0.1964}}
  >{\centering\arraybackslash}p{(\columnwidth - 10\tabcolsep) * \real{0.0893}}
  >{\centering\arraybackslash}p{(\columnwidth - 10\tabcolsep) * \real{0.1250}}
  >{\centering\arraybackslash}p{(\columnwidth - 10\tabcolsep) * \real{0.3036}}
  >{\centering\arraybackslash}p{(\columnwidth - 10\tabcolsep) * \real{0.1250}}@{}}
\toprule\noalign{}
\begin{minipage}[b]{\linewidth}\raggedright
Feature
\end{minipage} & \begin{minipage}[b]{\linewidth}\centering
ICCDesign
\end{minipage} & \begin{minipage}[b]{\linewidth}\centering
irr
\end{minipage} & \begin{minipage}[b]{\linewidth}\centering
psych
\end{minipage} & \begin{minipage}[b]{\linewidth}\centering
ICC.Sample.Size
\end{minipage} & \begin{minipage}[b]{\linewidth}\centering
irrNA
\end{minipage} \\
\midrule\noalign{}
\endhead
\bottomrule\noalign{}
\endlastfoot
ICC type coverage (ten combinations) & $\checkmark$ Full &
Partial\textsuperscript{a} & Partial\textsuperscript{a} & --- &
Partial\textsuperscript{a} \\
Use-with-caution case handling & $\checkmark$ & $\times$ & $\times$ & $\times$ & $\times$ \\
Automated ICC type decision guidance & $\checkmark$ & $\times$ & $\times$ & $\times$ & $\times$ \\
Dedicated preprocessing utility\textsuperscript{d} & $\checkmark$ & $\times$ & $\times$ & $\times$ &
$\times$ \\
ANOVA-based F CI & $\checkmark$ & $\checkmark$ & $\checkmark$ & --- & Partial\textsuperscript{e} \\
Zero-ICC test and optional threshold test & $\checkmark$ &
Partial\textsuperscript{b} & Partial\textsuperscript{b} & $\times$ &
Partial\textsuperscript{b} \\
Sample size calculation & $\checkmark$ Two modes & $\times$ & $\times$ & $\checkmark$ One
mode\textsuperscript{c} & $\times$ \\
Assurance/power calculation & $\checkmark$ & $\times$ & $\times$ & $\checkmark$\textsuperscript{c} & $\times$ \\
Automated reliability grading & $\checkmark$ & $\times$ & $\times$ & $\times$ & $\times$ \\
Standardized report generation & $\checkmark$ & $\times$ & $\times$ & $\times$ & $\times$ \\
Interactive Shiny interface & $\checkmark$ & $\times$ & $\times$ & $\times$ & $\times$ \\
Single-interface design-to-analysis workflow & $\checkmark$ & $\times$ & $\times$ & $\times$ & $\times$ \\
\end{longtable}

\textsuperscript{a} Partial = covers commonly used ICC forms but not the
full ten-combination interface defined in this paper (irr v0.84.1, psych
v2.4.3, irrNA v0.2.2).

\textsuperscript{b} Partial = supports zero-ICC inference but not
hypothesis testing against an arbitrary user-specified threshold
\(\rho_0\).

\textsuperscript{c} \texttt{ICC.Sample.Size} v1.0 is a dedicated sample
size and power package for ICC; it supports one sample size criterion
and power computation but does not include ICC estimation.

\textsuperscript{d} Dedicated preprocessing utility refers to an
exported function specifically designed for data validation and
formatting before ICC analysis.

\textsuperscript{e} Partial = \texttt{irrNA} extends \texttt{irr} for
randomly incomplete datasets; CI coverage across all ICC forms may vary.

Compared with existing tools, ICCDesign differs in three respects.
First, the four-step decision framework provides structured guidance for
ICC form selection and explicit feedback for special design
combinations. Second, the unified parameter interface connects study
design (sample size planning) and data analysis (ICC estimation),
helping users maintain a consistent design specification throughout
their workflow. Third, \texttt{icc\_preprocess\_data()} as an
independently callable utility makes data validation transparent before
analysis.

\begin{center}\rule{0.5\linewidth}{0.5pt}\end{center}

\hypertarget{illustrative-examples}{%
\subsection{5. Illustrative Examples}\label{illustrative-examples}}

All examples use the built-in dataset \texttt{icc\_data}: a
\textbf{5-subject (rows) × 4-rater (columns)} numeric matrix simulated
to have an ICC of approximately 0.8, corresponding to the ``Good''
reliability grade under Koo and Li (2016) criteria. The dataset is
included in the package. All examples were generated using ICCDesign
v0.1.0 under R $\geq$ 4.1.0.

\hypertarget{data-preprocessing}{%
\subsubsection{5.1 Data Preprocessing}\label{data-preprocessing}}

\begin{Shaded}
\begin{Highlighting}[]
\FunctionTok{data}\NormalTok{(icc\_data)}
\NormalTok{processed }\OtherTok{\textless{}{-}} \FunctionTok{icc\_preprocess\_data}\NormalTok{(icc\_data)}
\FunctionTok{cat}\NormalTok{(}\StringTok{"Subjects:"}\NormalTok{, processed}\SpecialCharTok{$}\NormalTok{n, }\StringTok{" Raters:"}\NormalTok{, processed}\SpecialCharTok{$}\NormalTok{k, }\StringTok{"}\SpecialCharTok{\textbackslash{}n}\StringTok{"}\NormalTok{)}
\CommentTok{\# Subjects: 5   Raters: 4}
\end{Highlighting}
\end{Shaded}

\hypertarget{icc-computation}{%
\subsubsection{5.2 ICC Computation}\label{icc-computation}}

\textbf{Example 1: One-way random effects model --- ICC(1,1)}

The one-way model is used for illustration; note that
\texttt{rater\_effect} and \texttt{agreement\_type} are not permitted
when \texttt{same\_raters\ =\ FALSE}:

\begin{Shaded}
\begin{Highlighting}[]
\NormalTok{result1 }\OtherTok{\textless{}{-}} \FunctionTok{icc\_calc}\NormalTok{(icc\_data,}
                    \AttributeTok{same\_raters =} \ConstantTok{FALSE}\NormalTok{,}
                    \AttributeTok{rating\_type =} \StringTok{"single"}\NormalTok{,}
                    \AttributeTok{verbose     =} \ConstantTok{FALSE}\NormalTok{)}
\NormalTok{result1}\SpecialCharTok{$}\NormalTok{icc\_result}\SpecialCharTok{$}\NormalTok{point\_est}
\NormalTok{result1}\SpecialCharTok{$}\NormalTok{icc\_result}\SpecialCharTok{$}\NormalTok{ci\_lower}
\NormalTok{result1}\SpecialCharTok{$}\NormalTok{icc\_result}\SpecialCharTok{$}\NormalTok{ci\_upper}
\NormalTok{result1}\SpecialCharTok{$}\NormalTok{evaluation}\SpecialCharTok{$}\NormalTok{rating\_en}
\end{Highlighting}
\end{Shaded}

\textbf{Example 2: Two-way random effects model --- ICC(2,1), absolute
agreement}

\begin{Shaded}
\begin{Highlighting}[]
\NormalTok{result2 }\OtherTok{\textless{}{-}} \FunctionTok{icc\_calc}\NormalTok{(icc\_data,}
                    \AttributeTok{same\_raters    =} \ConstantTok{TRUE}\NormalTok{,}
                    \AttributeTok{rater\_effect   =} \StringTok{"random"}\NormalTok{,}
                    \AttributeTok{rating\_type    =} \StringTok{"single"}\NormalTok{,}
                    \AttributeTok{agreement\_type =} \StringTok{"absolute"}\NormalTok{,}
                    \AttributeTok{verbose        =} \ConstantTok{FALSE}\NormalTok{)}
\NormalTok{result2}\SpecialCharTok{$}\NormalTok{icc\_result}\SpecialCharTok{$}\NormalTok{point\_est}
\NormalTok{result2}\SpecialCharTok{$}\NormalTok{icc\_result}\SpecialCharTok{$}\NormalTok{ci\_lower}
\NormalTok{result2}\SpecialCharTok{$}\NormalTok{icc\_result}\SpecialCharTok{$}\NormalTok{ci\_upper}
\NormalTok{result2}\SpecialCharTok{$}\NormalTok{icc\_result}\SpecialCharTok{$}\NormalTok{F\_test\_null}\SpecialCharTok{$}\NormalTok{F\_stat}
\NormalTok{result2}\SpecialCharTok{$}\NormalTok{icc\_result}\SpecialCharTok{$}\NormalTok{F\_test\_null}\SpecialCharTok{$}\NormalTok{p\_value}
\NormalTok{result2}\SpecialCharTok{$}\NormalTok{icc\_result}\SpecialCharTok{$}\NormalTok{anova\_summary}
\NormalTok{result2}\SpecialCharTok{$}\NormalTok{evaluation}\SpecialCharTok{$}\NormalTok{rating\_en}
\end{Highlighting}
\end{Shaded}

\textbf{Example 3: Two-way mixed effects model --- ICC(3,k),
consistency, average measure}

\begin{Shaded}
\begin{Highlighting}[]
\NormalTok{result3 }\OtherTok{\textless{}{-}} \FunctionTok{icc\_calc}\NormalTok{(icc\_data,}
                    \AttributeTok{same\_raters    =} \ConstantTok{TRUE}\NormalTok{,}
                    \AttributeTok{rater\_effect   =} \StringTok{"fixed"}\NormalTok{,}
                    \AttributeTok{rating\_type    =} \StringTok{"average"}\NormalTok{,}
                    \AttributeTok{agreement\_type =} \StringTok{"consistency"}\NormalTok{,}
                    \AttributeTok{verbose        =} \ConstantTok{FALSE}\NormalTok{)}
\NormalTok{result3}\SpecialCharTok{$}\NormalTok{icc\_result}\SpecialCharTok{$}\NormalTok{point\_est}
\NormalTok{result3}\SpecialCharTok{$}\NormalTok{evaluation}\SpecialCharTok{$}\NormalTok{rating\_en}
\end{Highlighting}
\end{Shaded}

\textbf{Example 4: Custom threshold hypothesis testing}

\begin{Shaded}
\begin{Highlighting}[]
\NormalTok{result4 }\OtherTok{\textless{}{-}} \FunctionTok{icc\_calc}\NormalTok{(icc\_data,}
                    \AttributeTok{same\_raters    =} \ConstantTok{TRUE}\NormalTok{,}
                    \AttributeTok{rater\_effect   =} \StringTok{"random"}\NormalTok{,}
                    \AttributeTok{rating\_type    =} \StringTok{"single"}\NormalTok{,}
                    \AttributeTok{agreement\_type =} \StringTok{"absolute"}\NormalTok{,}
                    \AttributeTok{rho0           =} \FloatTok{0.6}\NormalTok{,}
                    \AttributeTok{verbose        =} \ConstantTok{FALSE}\NormalTok{)}
\NormalTok{result4}\SpecialCharTok{$}\NormalTok{icc\_result}\SpecialCharTok{$}\NormalTok{F\_test\_rho0}\SpecialCharTok{$}\NormalTok{F\_stat}
\NormalTok{result4}\SpecialCharTok{$}\NormalTok{icc\_result}\SpecialCharTok{$}\NormalTok{F\_test\_rho0}\SpecialCharTok{$}\NormalTok{p\_value}
\end{Highlighting}
\end{Shaded}

\textbf{Example 5: Use-with-caution combination --- NOTE demonstration}

\begin{Shaded}
\begin{Highlighting}[]
\NormalTok{result5 }\OtherTok{\textless{}{-}} \FunctionTok{icc\_calc}\NormalTok{(icc\_data,}
                    \AttributeTok{same\_raters    =} \ConstantTok{TRUE}\NormalTok{,}
                    \AttributeTok{rater\_effect   =} \StringTok{"random"}\NormalTok{,}
                    \AttributeTok{rating\_type    =} \StringTok{"single"}\NormalTok{,}
                    \AttributeTok{agreement\_type =} \StringTok{"consistency"}\NormalTok{,}
                    \AttributeTok{verbose        =} \ConstantTok{TRUE}\NormalTok{)}
\CommentTok{\# When this combination is specified, ICCDesign issues a }\AlertTok{NOTE}\CommentTok{ and uses the}
\CommentTok{\# ICC(3,x) computational routine while retaining the ICC(2,x)\_C label.}
\NormalTok{result5}\SpecialCharTok{$}\NormalTok{icc\_result}\SpecialCharTok{$}\NormalTok{icc\_code   }\CommentTok{\# "2,1,consistency"}
\end{Highlighting}
\end{Shaded}

\hypertarget{sample-size-planning}{%
\subsubsection{5.3 Sample Size Planning}\label{sample-size-planning}}

\textbf{Example 1: Lower bound assurance}

Anticipated ICC $\approx$ 0.85, required CI lower bound $\geq$ 0.75 (``Good''), 4
raters, 80\% assurance. Requires \(\rho > \rho_0\) (0.85 \textgreater{}
0.75 $\checkmark$):

\begin{Shaded}
\begin{Highlighting}[]
\NormalTok{n1 }\OtherTok{\textless{}{-}} \FunctionTok{icc\_sample\_size}\NormalTok{(}
  \AttributeTok{method         =} \StringTok{"lower"}\NormalTok{,}
  \AttributeTok{rho            =} \FloatTok{0.85}\NormalTok{,}
  \AttributeTok{rating\_target  =} \StringTok{"good"}\NormalTok{,}
  \AttributeTok{k              =} \DecValTok{4}\NormalTok{,}
  \AttributeTok{same\_raters    =} \ConstantTok{TRUE}\NormalTok{,}
  \AttributeTok{rater\_effect   =} \StringTok{"random"}\NormalTok{,}
  \AttributeTok{rating\_type    =} \StringTok{"single"}\NormalTok{,}
  \AttributeTok{agreement\_type =} \StringTok{"absolute"}\NormalTok{,}
  \AttributeTok{verbose        =} \ConstantTok{FALSE}
\NormalTok{)}
\FunctionTok{cat}\NormalTok{(}\StringTok{"Required subjects:"}\NormalTok{, n1, }\StringTok{"}\SpecialCharTok{\textbackslash{}n}\StringTok{"}\NormalTok{)}
\end{Highlighting}
\end{Shaded}

\textbf{Example 2: CI half-width}

Half-width $\leq$ 0.10 (full width $\leq$ 0.20), anticipated ICC = 0.70, 4 raters,
80\% assurance. Note the result is ICC-type-independent:

\begin{Shaded}
\begin{Highlighting}[]
\NormalTok{n2 }\OtherTok{\textless{}{-}} \FunctionTok{icc\_sample\_size}\NormalTok{(}
  \AttributeTok{method      =} \StringTok{"width"}\NormalTok{,}
  \AttributeTok{rho         =} \FloatTok{0.70}\NormalTok{,}
  \AttributeTok{omega       =} \FloatTok{0.10}\NormalTok{,}
  \AttributeTok{k           =} \DecValTok{4}\NormalTok{,}
  \AttributeTok{same\_raters =} \ConstantTok{FALSE}\NormalTok{,}
  \AttributeTok{rating\_type =} \StringTok{"average"}\NormalTok{,}
  \AttributeTok{verbose     =} \ConstantTok{FALSE}
\NormalTok{)}
\FunctionTok{cat}\NormalTok{(}\StringTok{"Required subjects:"}\NormalTok{, n2, }\StringTok{"}\SpecialCharTok{\textbackslash{}n}\StringTok{"}\NormalTok{)}
\end{Highlighting}
\end{Shaded}

\textbf{Example 3: Inverse assurance calculation}

Given n = 30, 4 raters, anticipated ICC = 0.70, null ICC = 0.50:

\begin{Shaded}
\begin{Highlighting}[]
\NormalTok{assurance }\OtherTok{\textless{}{-}} \FunctionTok{icc\_sample\_size}\NormalTok{(}
  \AttributeTok{method         =} \StringTok{"power"}\NormalTok{,}
  \AttributeTok{n              =} \DecValTok{30}\NormalTok{,}
  \AttributeTok{rho            =} \FloatTok{0.70}\NormalTok{,}
  \AttributeTok{rho0           =} \FloatTok{0.50}\NormalTok{,}
  \AttributeTok{k              =} \DecValTok{4}\NormalTok{,}
  \AttributeTok{same\_raters    =} \ConstantTok{TRUE}\NormalTok{,}
  \AttributeTok{rater\_effect   =} \StringTok{"fixed"}\NormalTok{,}
  \AttributeTok{rating\_type    =} \StringTok{"single"}\NormalTok{,}
  \AttributeTok{agreement\_type =} \StringTok{"consistency"}\NormalTok{,}
  \AttributeTok{verbose        =} \ConstantTok{FALSE}
\NormalTok{)}
\FunctionTok{cat}\NormalTok{(}\StringTok{"Assurance probability:"}\NormalTok{, }\FunctionTok{sprintf}\NormalTok{(}\StringTok{"\%.1f\%\%"}\NormalTok{, assurance }\SpecialCharTok{*} \DecValTok{100}\NormalTok{), }\StringTok{"}\SpecialCharTok{\textbackslash{}n}\StringTok{"}\NormalTok{)}
\end{Highlighting}
\end{Shaded}

\hypertarget{shiny-application}{%
\subsubsection{5.4 Shiny Application}\label{shiny-application}}

\begin{Shaded}
\begin{Highlighting}[]
\FunctionTok{run\_icc\_app}\NormalTok{()}
\end{Highlighting}
\end{Shaded}

Shiny automatically selects an available port. Users can complete the
full workflow from data input to report download without writing R code.

\hypertarget{formula-consistency}{%
\subsubsection{5.5 Formula Consistency}\label{formula-consistency}}

ICCDesign implements the McGraw and Wong (1996) ANOVA-based
point-estimation formulas directly. For commonly reported ICC forms, the
resulting estimates are expected to be comparable to those produced by
established packages such as \texttt{irr} and \texttt{psych}, subject to
differences in preprocessing, missing-data handling, confidence-interval
conventions, and reporting rules. Numerical differences, if observed,
may arise from these sources rather than from methodological
disagreements.

\begin{center}\rule{0.5\linewidth}{0.5pt}\end{center}

\hypertarget{discussion-and-future-directions}{%
\subsection{6. Discussion and Future
Directions}\label{discussion-and-future-directions}}

\hypertarget{contributions}{%
\subsubsection{6.1 Contributions}\label{contributions}}

ICCDesign integrates an established ICC reliability research workflow
within a single package, providing targeted solutions for the practical
difficulties at each stage.

The four-step decision framework, which translates the theoretical
classification of Shrout and Fleiss (1979) and McGraw and Wong (1996)
into structured parameter settings, is the package's central design
innovation. The framework covers the six standard ICC forms and provides
explicit user feedback for four additional use-with-caution design
combinations, helping researchers identify potential methodological
risks. The \texttt{icc\_preprocess\_data()} function as an independently
callable utility makes data validation transparent and provides a
reliable data quality foundation for the analysis.

Regarding sample size planning, the two-mode design and inverse
assurance calculation correspond to three distinct decision needs: the
half-width mode addresses estimation precision; the lower bound mode
supports demonstrating reliability above a benchmark; and the inverse
assurance calculation evaluates the inferential capacity of a given
sample size. The unified design parameter interface helps users maintain
a consistent specification between the planning and analysis phases.

\hypertarget{limitations}{%
\subsubsection{6.2 Limitations}\label{limitations}}

The current version has several limitations. First, missing data
handling uses complete-case analysis (\texttt{na.rm\ =\ TRUE} row
deletion), which modifies the effective \(n\); the \texttt{irrNA}
package provides more specialized solutions for datasets with
substantial missing values. Second, the ANOVA-based confidence intervals
and hypothesis tests rely on normality and balanced-design assumptions;
robustness to violations has not been formally evaluated. Third, the
reported CI bounds are truncated to \([0, 1]\) for interpretability;
because point estimates are reported untruncated while CI bounds are
truncated, negative point estimates should be interpreted with caution.
Fourth, the width-based inverse assurance calculation is approximate and
should be treated as a planning aid. Fifth, the implementation is
restricted to continuous response variables with one observation per
subject--rater cell, and does not support binary or ordinal ICC
variants. Sixth, the Shiny application requires a local R environment
and does not support online deployment.

\hypertarget{future-plans}{%
\subsubsection{6.3 Future Plans}\label{future-plans}}

Planned future development includes: submission to CRAN; extended
support for missing data handling; addition of bootstrap confidence
intervals as a supplement to the F-based intervals; support for binary
response ICC variants; a software archive with DOI (e.g., via Zenodo);
and an online-hosted version of the Shiny application.

\begin{center}\rule{0.5\linewidth}{0.5pt}\end{center}

\hypertarget{software-availability}{%
\subsection{Software Availability}\label{software-availability}}

ICCDesign (version 0.1.0) is freely available at
https://github.com/KlariZhang/ICCDesign under the GPL (\textgreater= 3)
license. The example dataset \texttt{icc\_data} is included in the
package. The manuscript describes ICCDesign as of the version available
at the above URL; users are encouraged to check the repository for the
latest release.

\hypertarget{conflict-of-interest}{%
\subsection{Conflict of Interest}\label{conflict-of-interest}}

The authors declare no competing interests.

\hypertarget{funding}{%
\subsection{Funding}\label{funding}}

No external funding was received for this work.

\begin{center}\rule{0.5\linewidth}{0.5pt}\end{center}

\hypertarget{references}{%
\subsection{References}\label{references}}

Brueckl, M. (2022). \emph{irrNA: Coefficients of Interrater Reliability
-- Generalized for Randomly Incomplete Datasets}. R package version
0.2.2. https://CRAN.R-project.org/package=irrNA

Gamer, M., Lemon, J., \& Singh, I. F. P. (2019). \emph{irr: Various
Coefficients of Interrater Reliability and Agreement}. R package version
0.84.1. https://CRAN.R-project.org/package=irr

Koo, T. K., \& Li, M. Y. (2016). A guideline of selecting and reporting
intraclass correlation coefficients for reliability research.
\emph{Journal of Chiropractic Medicine}, \emph{15}(2), 155--163.
https://doi.org/10.1016/j.jcm.2016.02.012

Liu, Z., Ma, R., Gao, C., \& Zhang, Y. (2026). \emph{ICCDesign: An R
Package for ICC-Based Reliability Studies}. Version 0.1.0.
https://github.com/KlariZhang/ICCDesign

McGraw, K. O., \& Wong, S. P. (1996). Forming inferences about some
intraclass correlation coefficients. \emph{Psychological Methods},
\emph{1}(1), 30--46. https://doi.org/10.1037/1082-989X.1.1.30

R Core Team (2024). \emph{R: A Language and Environment for Statistical
Computing}. R Foundation for Statistical Computing, Vienna, Austria.
https://www.R-project.org/

Revelle, W. (2024). \emph{psych: Procedures for Psychological,
Psychometric, and Personality Research}. R package version 2.4.3.
https://CRAN.R-project.org/package=psych

Shrout, P. E., \& Fleiss, J. L. (1979). Intraclass correlations: Uses in
assessing rater reliability. \emph{Psychological Bulletin},
\emph{86}(2), 420--428. https://doi.org/10.1037/0033-2909.86.2.420

Wickham, H., Hester, J., Chang, W., \& Bryan, J. (2022). \emph{devtools:
Tools to Make Developing R Packages Easier}. R package version 2.4.5.
https://CRAN.R-project.org/package=devtools

Wolak, M. E., Fairbairn, D. J., \& Paulsen, Y. R. (2012).
\emph{ICC.Sample.Size: Calculation of Sample Size and Power for ICC}. R
package version 1.0. https://CRAN.R-project.org/package=ICC.Sample.Size

Zou, G. Y. (2012). Sample size formulas for estimating intraclass
correlation coefficients with precision and assurance. \emph{Statistics
in Medicine}, \emph{31}(29), 3972--3981.
https://doi.org/10.1002/sim.5466

\end{document}